\documentclass{aa}
\usepackage{graphicx}

\begin{document}

\newcommand{\lsim}{\mathrel{\rlap{\raise -.3ex\hbox{${\scriptstyle\sim}$}}%
                   \raise .6ex\hbox{${\scriptstyle <}$}}}%
\newcommand{\gsim}{\mathrel{\rlap{\raise -.3ex\hbox{${\scriptstyle\sim}$}}%
                   \raise .6ex\hbox{${\scriptstyle >}$}}}%

\title{Radio source contamination of the Sunyaev-Zeldovich effect in galaxy clusters}
\titlerunning{Radio source contamination of the Sunyaev-Zeldovich effect}
\author{M. Massardi\inst{1}
        \and
          G. De Zotti\inst{1,2}
          }

   \offprints{G. De Zotti: dezotti@pd.astro.it}

\institute{
   INAF--Osservatorio Astronomico di Padova, Vicolo dell'Osservatorio 5,
   I-35122 Padova,
   Italy\\
   \and International School for Advanced Studies,
SISSA/ISAS, Via Beirut 2-4, I-34014 Trieste, Italy\\
    }

\date{Received / Accepted }







\abstract{By cross-correlating the 1.4 GHz FIRST catalog of radio
sources with the Abell cluster catalog we have found an excess
surface density by a factor $\simeq 5$ of radio sources within a
projected distance, $r_p$, of 0.1 Mpc (for $h=0.65$) from the
cluster center. The profile of the excess density can be
described, for $r_p \gsim 0.1\,$Mpc, by a $\beta$-model  with a
core radius of $\simeq 0.70\,$Mpc and $\beta=1.65$. The luminosity
function of cluster sources does not show hints of cosmological
evolution over the redshift range ($z\lsim 0.4$) covered by our
cluster sample. The mean luminosity function is in excellent
agreement with the recent determination by Reddy \& Yun (2004) for
7 nearby clusters and extends it by two orders of magnitude to
higher luminosities. Its shape is very similar to that of the
local luminosity function of field galaxies, but the space density
is about 3000 times higher. When extrapolated to 30 GHz, our
luminosity function compares very favourably with an estimate
obtained directly from the 30 GHz observations by Cooray et al.
(1998). The antenna temperature contributed by radio sources
within the nominal cluster radius of $1.7\,$Mpc is estimated to be
$\simeq 13.5\,\mu$K at 30 GHz for clusters at $z\simeq 0$ and
decreases to $\simeq 3.4\,\mu$K at $z\simeq 0.5$, in the absence
of cosmological evolution; it increases by a factor of $\simeq
1.5$ within 0.25 Mpc from the cluster center. If the pure
luminosity evolution models by Dunlop \& Peacock (1990) are
adopted, the radio source antenna temperature turns out to be
essentially independent of redshift.}

\maketitle






\section{Introduction}
Over the last decade the quality of observations of the
Sunyaev-Zeldovich (SZ) effect (Sunyaev \& Zeldovich 1972, 1980,
1981) towards galaxy clusters has dramatically improved, so that
it is now routinely detected with high S/N and even imaged
(Carlstrom et al. 2002). To fully exploit the extraordinary
information content of the effect, however, one needs to take into
account the sources of possible contamination.

Since powerful radio sources are normally associated to early-type
galaxies and these preferentially reside in clusters, a strong
over-density of radio sources, which could partly fill the SZ dip
in the Rayleigh-Jeans region, is naturally expected in clusters.
And, indeed, radio sources were found to be the major contaminant
of experiments using centimeter-wavelength receivers (Cooray et
al. 1998; LaRoque et al. 2003).

In this paper, we exploit the FIRST (Becker et al. 1995; White et
al. 1997) survey at 1.4 GHz to investigate the correlation
function of radio sources with Abell clusters (Abell 1958) and to
derive an estimated of the luminosity function of radio sources in
clusters. Coupling the FIRST with the GB6 (Gregory et al. 1996)
survey, we obtain an estimate of the spectral index distribution
that we use to extrapolate the luminosity function to centimeter
wavelengths and to estimate the mean contamination of the SZ
signal as a function of the cluster redshift.

\section{FIRST sources within cluster solid angles}

\subsection{Source selection}

We have used the April 2003 update of the FIRST survey catalog
(sundog.stsci.edu/first/catalogs.html), containing 811,117 radio
sources over an area of 8,422 square degrees in the North Galactic
cap and 611 square degrees in the South Galactic cap, with a
detection threshold of about 1 mJy and an angular resolution of
$5''$. The integrated flux was adopted for sources labelled as
extended and the peak flux for the others.

We have looked for FIRST sources within a projected distance of
$3\,$Mpc (typical cluster virial radius), $1\,$Mpc, and
$0.25\,$Mpc from the Abell cluster centers, for a flat
$\Lambda$CDM cosmology ($\Omega_m=0.3$, $\Omega_\Lambda=0.7$) and
$H_0= 65\,\hbox{km}\,\hbox{s}^{-1}\,\hbox{Mpc}^{-1}$ (or
$h=H_0/100=0.65$). We have kept only those clusters fully within
the FIRST area (taking into account the ``holes'' in such area).
We have also excluded clusters partially overlapping with each
other.

Measured redshifts for 368 Abell clusters in the FIRST area are
listed in Struble \& Rood (1999); 1 additional redshift
measurement was found in Pimbblet et al. (2001). Redshift
estimates for the remaining 582 clusters were obtained using the
relationship between redshift and magnitude of the tenth brightest
cluster galaxy derived by Postman et al. (1985), which includes a
correction for the Scott effect. A comparison of estimated with
measured redshift showed this relationship to be superior to those
proposed by other Authors (Fullerton et al. 1972; Leir et al.
1974; Kalinkov et al. 1994).

Our final sample comprises 951 clusters with a total of 617,
5,574, and 35,526 radio sources within 0.25 Mpc, 1 Mpc, and 3 Mpc
from the cluster centers, respectively. We have also selected 738
control fields, each having an angular radius of $0^\circ.5$ and
whose centers are more than $1^\circ$ away from the center of any
of the clusters. The control fields contain a total of 51,770
radio sources.

\begin{table*}
\caption[]{Mean surface densities ($\hbox{sr}^{-1}$) of FIRST
radio sources within projected distances of 0.25, 1 and 3 Mpc
($\bar{n}_{0.25}$, $\bar{n}_{1}$, and $\bar{n}_{3}$, respectively)
from cluster centers, for different redshift intervals (the
notation xxE+y stands for xx$\times 10^y$). $N_{\rm cl}$ is the
number of clusters in a given redshift bin. The average surface
density of control fields is $(2.93 \pm 0.545)\times
10^5\,\hbox{sr}^{-1}$.} \label{densities} \centering
\begin{tabular}{ccrccc}
  \hline
$\log z_{\rm min}$ & $\log z_{\rm max}$ & $N_{\rm cl}$ &
$\bar{n}_{0.25}$ & $\bar{n}_{1}$&$\bar{n}_{3}$\\ \hline
 -1.80 & -1.15 & 41 & $(7.7\pm 1.5)$E+5 & $(3.93\pm 0.27)$E+5 & $(2.96\pm 0.06)$E+5 \\
 -1.15 & -1.00 & 81 & $(7.8\pm 1.1)$E+5 & $(4.15\pm 0.24)$E+5 & $(3.03\pm 0.07)$E+5 \\
 -1.00 & -0.90 &130 & $(8.1\pm 1.1)$E+5 & $(4.13\pm 0.23)$E+5 & $(3.13\pm 0.07)$E+5 \\
 -0.90 & -0.85 &201 & $(7.7\pm 1.0)$E+5 & $(4.64\pm 0.22)$E+5 & $(3.16\pm 0.07)$E+5 \\
 -0.85 & -0.80 &149 & $(8.1\pm 1.4)$E+5 & $(4.62\pm 0.25)$E+4 & $(3.19\pm 0.07)$E+5 \\
 -0.80 & -0.75 &168 & $(7.8\pm 1.3)$E+5 & $(5.12\pm 0.26)$E+5 & $(3.24\pm 0.08)$E+5 \\
 -0.75 & -0.70 &105 & $(7.9\pm 1.5)$E+5 & $(5.02\pm 0.35)$E+5 & $(3.29\pm 0.11)$E+5 \\
 -0.70 & -0.35 & 76 & $(1.4\pm 0.4)$E+6 & $(7.10\pm 0.61)$E+5 & $(3.76\pm 0.15)$E+5 \\
 \hline
\end{tabular}
\end{table*}

\begin{figure}[htbp]
\begin{center}
\includegraphics[height=8cm]{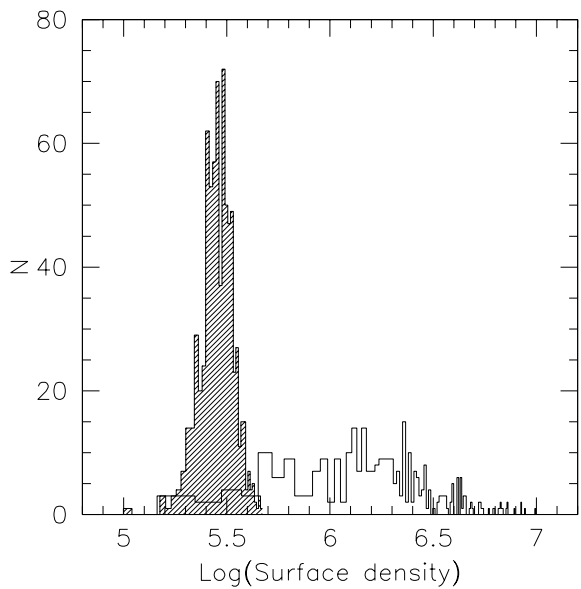}
\caption{Surface density (sr$^{-1}$) distribution of radio sources
in the control fields (shaded area) and for clusters in the lowest
redshift interval, within 0.25 Mpc from their centers.}
\label{fig:surfacedens}
\end{center}
\end{figure}

\begin{figure}[htbp]
\begin{center}
\includegraphics[height=8cm]{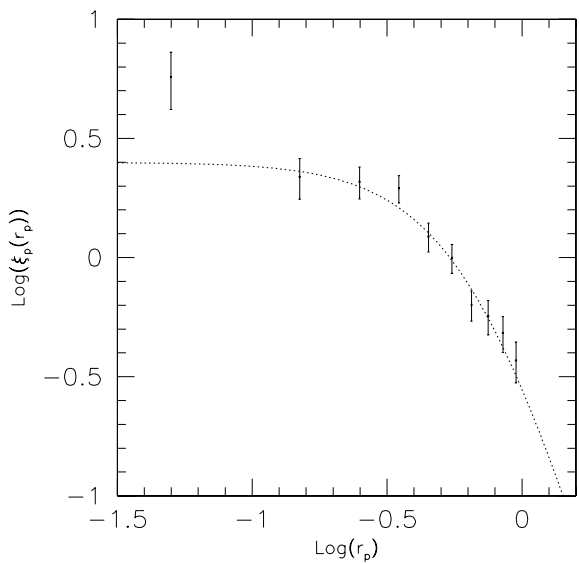}
\caption{Projected two-point correlation function of radio sources
with cluster centers, as a function of project radius (Mpc, for
$h=0.65)$. The dotted line shows a $\beta$-model fit with
$r_0=0.70\,$Mpc and $\beta=1.65$ [see Eq.~(\protect\ref{beta})].}
\label{fig:xi}
\end{center}
\end{figure}

\subsection{Analysis of the sample}

The cluster sample has been divided into 8 redshift intervals. The
mean surface densities within 0.25, 1, and 3 Mpc from the cluster
centers, and their errors are listed, for each interval, in
Table~\ref{densities}. Figure~\ref{fig:surfacedens} shows an
example of the distribution of surface densities within 0.25 Mpc,
$n_{0.25}$, compared with the distribution for control fields. The
mean value of $n_{0.25}$ ($\bar{n}_{0.25}$) is typically almost a
factor of 3 higher than in control fields, without any significant
trend with redshift. Cooray et al. (1998), who surveyed 56
clusters, within $250''$ from the pointing centers, at 28.5 GHz,
to a $4\sigma$ detection limit $\sim 0.4\,$mJy, found 4 to 7 times
more sources than expected from a simple extrapolation of
low-frequency counts outside clusters.

The over-density towards clusters is still quite substantial
within 1 Mpc from the cluster centers and is weak, but still
statistically significant, within 3 Mpc. A Kolmogorov-Smirnov test
gives, for the first redshift bin in Table~\ref{densities},
probabilities of $\simeq 2.8\times 10^{-12}$ and of $\simeq
6.5\times 10^{-8}$ that the distributions of $n_{0.25}$ and of
$n_{1}$, respectively, are drawn from the same parent distribution
as for control fields. Such probability drops by orders of
magnitude for the higher redshift bins. The difference with
control fields of the distribution of $n_{3}$ is not significant
for the lowest redshift bin, is $\simeq 1.6\times 10^{-2}$, for
the second lowest redshift, is $\simeq 2\times 10^{-3}$ for the
sixth bin, and is orders of magnitude lower for all the other
bins. The densities ${n}_{1}$ and ${n}_{3}$ correlate with
redshift, although the redshift dependence is rather weak.

To characterize more quantitatively the over-density, we have
computed the projected radio source-cluster two-point correlation
function, $\xi_p$, as a function of the projected linear distance,
$r_p$, from the cluster center, using the formula:
\begin{equation}
\xi_p(r_p)+1= \frac{N_c(r_p)}{N_f(r_p)}\ , \label{xi}
\end{equation}
where $N_c$ is the mean number of sources within annuli of width
$\Delta r = 0.1\,$Mpc, centered in the cluster centers, $N_f$ is
the mean number of control field sources in areas equal to those
of the annuli, and $r_p$ is equal to the inner radius of annuli
plus $\Delta r/2$. The errors on $\xi_p(r_p)$ where estimated as
$\delta \xi_p(r_p)\simeq (1+\xi_p(r_p))/(\sqrt{N_r})$ [Peacock
1999, Eq. (16.111)].

As shown by Fig.~\ref{fig:xi}, the surface density of sources
close to the cluster center is about a factor of 5 higher than in
the control fields, and decreases by more than a factor of 10 at a
radius of $\simeq 1\,$Mpc. Except for the innermost point, the
projected correlation function, $\xi_p(r_p)$ can be described by a
$\beta$-model (Cavaliere \& Fusco-Femiano 1976) surface-density
profile
\begin{equation}
\sigma(r_p)\propto \left(1+{r_p^2\over
r_0^2}\right)^{-(3/2)\beta+1/2} \label{beta}
\end{equation}
with $r_0=0.70\,$Mpc and $\beta =1.65$. For comparison, Reddy \&
Yun (2004) found, for a sample of 182 radio galaxies in 7 nearby,
rich, X-ray luminous clusters, $r_0=0.74\,$Mpc (scaled to our
choice for the Hubble constant) and $\beta=1.05$.


\begin{figure}[htbp]
\begin{center}
\includegraphics[height=8cm]{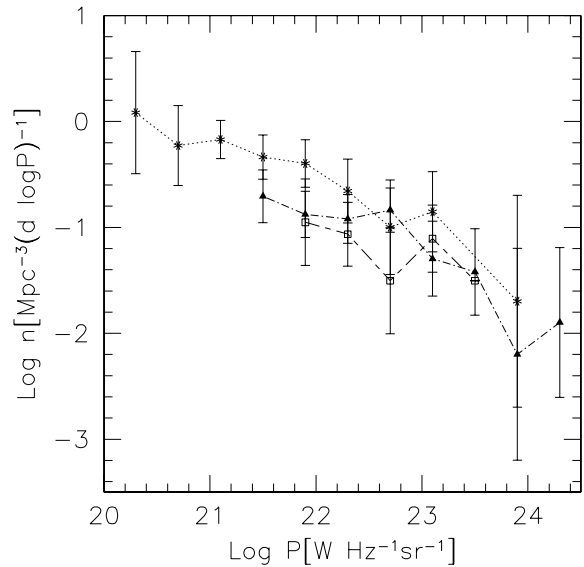}
\caption{Estimated luminosity functions in three representative
redshift bins: the first (stars), the third (filled triangles),
and the seventh (open squares) in Table~\protect\ref{densities}.
 }
\label{fig:LFz}
\end{center}

\end{figure}

\begin{figure}[htbp]
\begin{center}
\includegraphics[height=8cm]{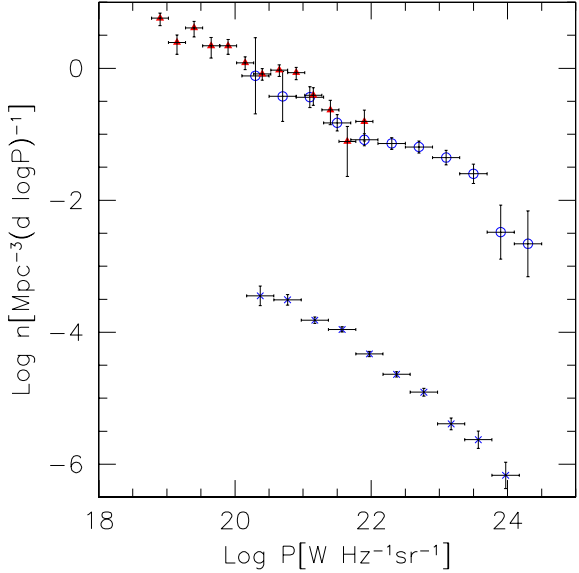}
\caption{Present estimate of the mean 1.4 GHz luminosity function
in clusters (open circles) compared with that by Reddy \& Yun
(2004; filled triangles) and with the local luminosity function in
the field (Magliocchetti et al. 2002; crosses). } \label{fig:LF}
\end{center}
\end{figure}

\begin{figure}[htbp]
\begin{center}
\includegraphics[height=8cm]{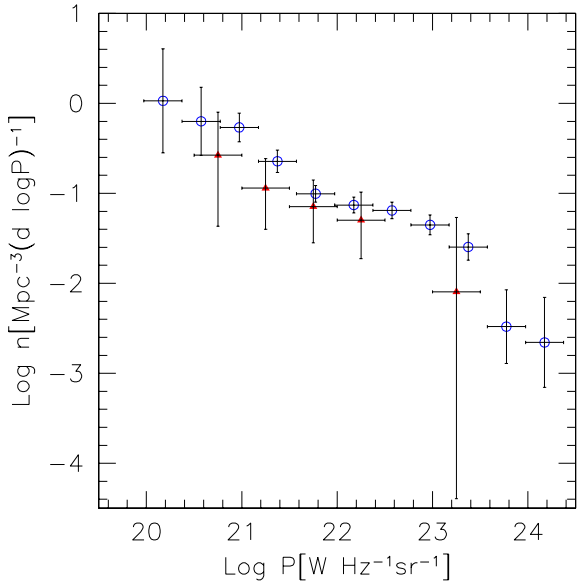}
\caption{Estimates of the luminosity function of cluster radio
galaxies at 30 GHz. The open circles show the extrapolation to 30
GHz of the luminosity function in Fig.~\protect{\ref{fig:LF}}. The
filled triangles represent our estimate based on the 30 GHz data
by Cooray et al. (1998).  } \label{fig:LFcooray}
\end{center}
\end{figure}

\begin{table}
\begin{center}
\caption{Estimate of the 1.4 GHz mean luminosity function of
cluster radio sources.}\label{tab:flum}
\begin{tabular}{cc}
  \hline
$\log[L(\hbox{W}\,\hbox{Hz}^{-1}\,\hbox{sr}^{-1})]$& $\log[n(\hbox{Mpc}^{-3}\,d\log L^{-1})]$\\
\hline
20.3    &   $\phantom{-}0.12 \pm  0.58$   \\
20.7    &   $\phantom{-}0.43 \pm  0.38$   \\
21.1    &   $\phantom{-}0.44 \pm  0.16$   \\
21.5    &   $-0.83  \pm   0.12$   \\
21.9    &   $-1.08  \pm   0.09$   \\
22.3    &   $-1.14  \pm   0.09$   \\
22.7    &   $-1.19  \pm   0.09$   \\
23.1    &   $-1.35  \pm   0.11$   \\
23.5    &   $-1.60  \pm   0.15$   \\
23.9    &   $-2.48  \pm   0.41$   \\
24.3    &   $-2.66  \pm   0.50$   \\
 \hline
\end{tabular}
\end{center}
\end{table}

\section{Luminosity function of radio galaxies in clusters}

Reddy \& Yun (2004) have estimated the mean 1.4 GHz luminosity
function of galaxies in 7 nearby clusters containing a total of
182 NVSS (Condon et al. 1998) sources, assuming a uniform
distribution within a cluster radius of 1.5 Mpc for $h=0.75$ (or
1.73 Mpc for the value of the Hubble constant, $h=0.65$, adopted
here).

Since the median density of radio galaxies within 0.25 Mpc from
the cluster center is $\simeq 8\times 10^5$ sources/sr
(Table~\ref{densities}), i.e. a factor $\simeq 2.7$ larger than
the field density, an average of 63\% of the 617 galaxies within
such radius are cluster members and we may attempt to estimate
their luminosity function. Following Reddy \& Yun (2004), we have
assumed a uniform distribution within a cluster radius of 1.73 Mpc
(for our choice of the Hubble constant); the sampled volume is
then the intersection of such sphere with a cylinder with radius
of 0.25 Mpc. The derived densities have been scaled down by a
factor of 0.63 to take into account the foreground/background
contamination and by a further factor of 4.48 to convert the mean
density within the sampled volume into the mean density within the
adopted cluster radius.

To check for possible evolutionary effects we have carried out
separate estimates for each redshift bin in Table~\ref{densities}.
Results for three representative redshift bins are shown in
Fig.~\ref{fig:LFz}. No indication of evolution is apparent. The
low luminosity excess found for the lowest luminosity bin is
likely due to contamination. Since we consider sources within a
fixed {\it physical} radius the number (but not the fraction) of
outliers (proportional to the corresponding solid angle) increases
with decreasing redshift; the flux densities of field sources are
preferentially close to the survey limits and therefore affect
more the lowest luminosity bins.

In the absence of evidences of cosmological evolution of the
luminosity function, we have lumped all galaxies together to
derive the mean luminosity function presented in
Table~\ref{tab:flum}. Figure~\ref{fig:LF} shows that our estimate
nicely agrees with that by Reddy \& Yun (2004), based on totally
different approach, and extends it by two orders of magnitude in
luminosity. The figure also compares the cluster with the field
local luminosity function recently determined by Magliocchetti et
al. (2002). The shape of the luminosity functions is similar but
density of radio sources in clusters is about 3000 times higher
than in the field. For comparison, the matter overdensity in a
cluster formed at a redshift $z_f$ is $\simeq 200(1+z_f)^3\simeq
3000$ for a formation redshift $z_f \simeq 1.5$.

The integral of the 1.4 GHz cluster luminosity function, made
adopting the estimate by Reddy and Yun (2004) for $\log
L(\hbox{W}\,\hbox{Hz}^{-1}\,\hbox{sr}^{-1}) \le 22$ and our
results for higher luminosities, gives a mean cluster radio
luminosity $\log L_{{\rm cluster},1.4{\rm GHz}} \simeq 23.35$.

\begin{figure}[htbp]
\begin{center}
\includegraphics[height=8cm]{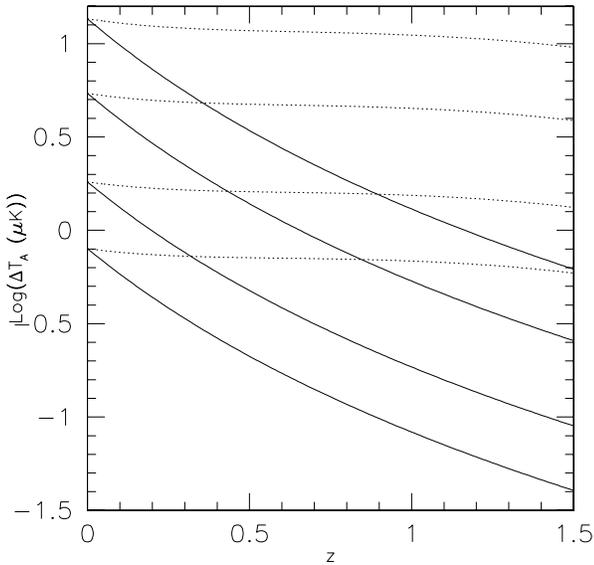}
\caption{Mean contamination of the SZ signal (in antenna
temperature) by radio sources as a function of cluster redshift
for 4 frequencies, 30, 44, 70, and 100 GHz (from top to bottom).
At each frequency, the solid line refers to the case of
no-evolution, the dotted line to the pure luminosity evolution
models for steep- and flat-spectrum sources described in Sect.
3.4.1 of Dunlop \& Peacock (1990).} \label{fig:cont}
\end{center}
\end{figure}

\begin{figure}[htbp]
\begin{center}
\includegraphics[height=8cm]{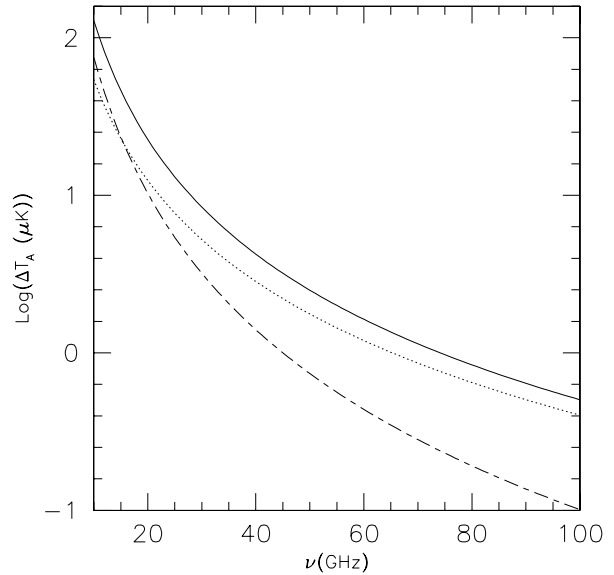}
\caption{Frequency dependence of the total emission (in terms of
antenna temperature) from cluster sources for $z=0.15$, the median
redshift of our cluster sample. The dashed and dotted lines
correspond to steep- and flat-spectrum sources, respectively,
while the solid line shows the total. } \label{fig:contnu}
\end{center}
\end{figure}

\section{Radio source contamination of Sunyaev-Zeldovich signals}

To extrapolate the 1.4 GHz cluster luminosity to the higher
frequencies used for measurements of the Sunyaev-Zeldovich effect,
we need the spectral index distribution of sources. To get some
information on that, we have cross-correlated the GB6 4.85 GHz
catalog (Gregory et al. 1996) with our cluster sample, looking for
sources within 0.25 Mpc from the cluster center. The number of
matches is relatively small (46 sources) because of the relatively
bright flux density limit (18 mJy, to be compared with the 1 mJy
FIRST limit). The identification of GB6 with FIRST sources is
complicated by the relatively poor GB6 angular resolution. In
fact, in several cases we have multiple FIRST sources within the
GB6 beam ($\hbox{FWHM} \simeq 3'.5$). In these cases, we have
taken as the 1.4 GHz flux density of the GB6 sources the sum of
FIRST flux densities weighted with a Gaussian response function
with $\hbox{FWHM} \simeq 3'.5$ and axis towards the nominal GB6
position. Assuming power-law spectra ($S\propto \nu^{-\alpha}$) we
found that, of the 46 GB6 sources, 12 are flat-spectrum ($\alpha <
0.5$) and 34 are steep-spectrum ($\alpha \ge 0.5$), so that the
ratio of flat to steep spectrum sources is 0.35($+0.065$,
$-0.063$), consistent with the ratio (0.41) we found for a random
sample of 1548 field sources. The mean spectral indices are
$\bar{\alpha}_{\rm flat} = 0.13$ and $\bar{\alpha}_{\rm steep} =
0.87$.

The ratio of 4.85 GHz luminosity densities of flat- to
steep-spectrum cluster sources is 0.38. Adopting the above mean
spectral indices, such ratio becomes 0.15 at 1.4 GHz. Using this
ratio, we partition the total cluster radio luminosity as $\log
L_{{\rm cluster,flat},1.4{\rm GHz}} \simeq 22.48$ and $\log
L_{{\rm cluster,steep},1.4{\rm GHz}} \simeq 23.29$ and extrapolate
these luminosities to higher frequencies using the corresponding
mean spectral indices.

The big extrapolation in frequency required to reach the spectral
range where most SZ surveys are carried out ($\nu \ge 20\,$GHz)
clearly makes our estimates, based on limited information on the
1.4 to 4.85 GHz spectral indices, liable to large uncertainties.
In fact, the low-frequency spectral indices may not apply up to
much higher frequencies. Also, sources with strongly inverted
spectra, such as extreme GHz Peaked Spectrum sources, undetected
by low frequency surveys, may become important at high
frequencies. An important test of our estimates is provided by the
30 GHz cluster survey by Cooray et al. (1998). These Authors have
detected at $\ge 4\sigma$ a total of 35 radio sources, towards 56
clusters. If indeed, as argued by Cooray et al. (1998), most
detected sources are physically associated to the clusters, we may
derive their luminosities attributing to them the redshift of the
associated cluster and estimate a mean 30 GHz luminosity function
of cluster sources. Extrapolating to 30 GHz the mean surface
densities of flat- and steep-spectrum sources in our control
fields, we find that the fraction of background/foreground sources
in their cluster fields is $\simeq 25\%$, consistent with the
estimates by Cooray et al. (1998). The derived 30 GHz luminosity
function was correspondingly scaled down by a factor of 1.25. In
Fig.~\ref{fig:LFcooray} this estimate is compared with the
extrapolation to 30 GHz of our 1.4 GHz luminosity function, using
the previously determined mean spectral indices for flat- and
steep-spectrum sources. Although, due to the smallness of the 30
GHz sample, the uncertainties are large and the region of overlap
is limited, the agreement is reassuringly good, supporting our
choices for the mean spectral indices and indicating that we are
not missing important source populations.

The contamination of the SZ effect by radio sources is more
usefully expressed in terms of antenna temperature as:
\begin{equation}
\Delta T_A(\nu,z)={L_{\rm flat}+L_{\rm steep}\over 2\pi k_b
\lambda^2 r_{\rm cluster}^2 (1+z)^3}, \label{eqn:Tant}
\end{equation}
where $L_{\rm flat(steep)} = L_{{\rm cluster,flat(steep)},1.4{\rm
GHz}}(\nu(1+z)$ $/1.4{\rm GHz})^{-\bar{\alpha}_{\rm
flat(steep)}}$, $k_b$ is the Boltzmann constant and $r_{\rm
cluster}$ is the physical radius of clusters, assumed here to be
1.73 Mpc.

The results are shown as a function of the cluster redshift, for 4
frequencies, in Fig.~\ref{fig:cont} where we have considered two
cases: no-evolution, as suggested by the analysis in
Fig.~\ref{fig:LFz} (solid lines), and pure luminosity evolution
(Dunlop \& Peacock 1990; dotted lines). The latter are probably to
be taken as upper limits, at least for the redshift range covered
by our cluster sample ($z \lsim 0.4$). The frequency dependence of
the antenna temperature for $z=0.15$, the median redshift of our
cluster sample, is shown in Fig.~\ref{fig:contnu}.

\section{Conclusions}

A cross-correlation of FIRST radio sources with Abell clusters has
demonstrated an excess surface density by a factor $\simeq 5$ of
radio sources within a projected distance, $r_p$, of 0.1 Mpc (for
$h=0.65$) from the cluster center. At larger radii, the profile of
the excess density can be described by a $\beta$-model (Cavaliere
\& Fusco-Femiano 1976) with a core radius of $\simeq 0.70\,$Mpc
and $\beta=1.65$ [see Eq.~(\ref{beta})].

Within $r_p=0.25\,$Mpc, chosen as a compromise between the
conflicting requirements of maximizing the sample size on one side
and of minimizing the foreground/background contamination on the
other side, we have a total of 617 sources within 951 Abell
clusters, at redshifts $\lsim 0.4$, with a median surface density
a factor of 2.7 higher than in the field, implying that 63\% of
sources are cluster members. Their mean luminosity function does
not show any evidence for cosmological evolution. Lumping all
sources together and taking into account the surface density
profile we have obtained an estimate of the mean luminosity
function at 1.4 GHz in excellent agreement with that derived by
Reddy \& Yun (2004) studying 7 nearby, X-ray luminous clusters, in
the overlapping luminosity range, and extending to luminosities
two orders of magnitude higher.

The lack of evidence for cosmological evolution up to $z \simeq
0.4$ is at odds with the predictions of the pure luminosity
evolution models by Dunlop \& Peacock (1990), indicating that
either the evolutionary behaviour of sources in clusters and in
the field differ or that such models overestimate the radio source
evolution at low redshifts.

No significant differences between the properties of sources in
clusters and in the field have emerged. The shape of our estimated
luminosity function is very similar to the local luminosity
function in the field at the same frequency (1.4 GHz), recently
determined by Magliocchetti et al. (2002). The mean density is
however about 3000 times higher, a factor that corresponds to the
matter overdensity in clusters formed at $z\simeq 1.5$.

The distribution of spectral indices, derived exploiting the 4.85
GHz GB6 survey, is also consistent with being the same in clusters
and in the field. Having estimated the fractions of steep- and
flat-spectrum sources and the mean spectral indices of the two
populations we have derived the mean emission of cluster sources,
as a function of redshift and of frequency. The 30 GHz luminosity
function obtained extrapolating our estimate at 1.4 GHz is found
to be in good agreement with that derived directly from the 30 GHz
observations by Cooray et al. (1998).

If sources do not evolve, their mean contamination of cluster SZ
effects, in terms of antenna temperature, decreases with
increasing redshift. For example, for observations at 30 GHz with
angular resolution matching the assumed cluster radius of 1.7 Mpc
($h=0.65$) we estimate a mean contamination of $\simeq 13.5\,\mu$K
at $z\simeq 0$ and of $\simeq 3.4\,\mu$K at $z\simeq 0.5$. The
contamination increases by a factor of 1.5 within 0.25 Mpc from
the cluster center. If the pure luminosity evolution models by
Dunlop \& Peacock (1990) are adopted, the contamination level
turns out to be very weakly redshift dependent. Obviously, the
contamination level drops rapidly with increasing frequency and
becomes pretty large at wavelengths above a few centimeters.

\begin{acknowledgements}

We thank the referee for having suggested the comparison with the
data by Cooray et al. (1998). Work supported in part by MIUR
(through a COFIN grant) and ASI.

\end{acknowledgements}

\end{document}